\documentclass[10pt,amstex]{article}
\usepackage{amssymb,amsmath,cite}
\usepackage{latexsym}
\numberwithin{equation}{section}

\usepackage{amsmath,amssymb}
\usepackage{graphicx}% Include figure files
\usepackage{dcolumn}% Align table columns on decimal point
\usepackage{bm}% bold math

\begin{document}
\begin{titlepage}
\begin{flushright}
September 15, 2011 \\
NSF-KITP-11-200 \\
ULB-TH/11-12
\end{flushright}
\vskip 1in

\begin{center}
{\Large{SUSY Enhancements in $(0,4)$ Deformations of $AdS_3$/CFT$_2$}}
\vskip 0.5in  St\'ephane Detournay$^{1}$, Joshua M. Lapan$^{2,3}$, and Mauricio 
%`Rockin'
 Romo$^{4}$
 \vskip 0.4in {\it $^{1}$ Physique Th\'eorique et Math\'ematique and International Solvay Institutes,\\ Universit\'e Libre de Bruxelles,\\  Campus Plaine, CP 231, Boulevard du triomphe \\ B-1150 Brussels, Belgium}
\vskip 0.1in {\it $^{2}$ Kavli Institute for Theoretical Physics, University of California,  Santa Barbara, CA}
\vskip 0.1in {\it $^3$ Department of Physics, McGill University, Montreal, QC, Canada}
\vskip 0.1in {\it $^{4}$ Department of Physics, University of California, Santa Barbara, CA}
\end{center}
\vskip 0.5in

\begin{abstract}

\noindent We discuss a marginal deformation of the $SL(2,R) \times SU(2) \times U(1)^4$ WZW model, which describes string theory on $AdS_3 \times S^3 \times T^4$, that corresponds to warping the $S^3$ factor.  This deformation breaks part of the $\mathcal{N}=(4,4)$ supersymmetry of the undeformed dual CFT to $\mathcal{N}=(0,4)$ supersymmetry.  In the spirit of work by Giveon, Kutasov, and Seiberg, we construct the asymptotic spacetime symmetry algebra from worldsheet operators and find a restoration of $(4,4)$ supersymmetry at discrete values of the deformation parameter.  We explain this result from various perspectives: the worldsheet, supergravity, and from the singular D1-D5 CFT. The supergravity analysis includes an asymptotic symmetry computation of the level of the affine $SU(2)$ R-symmetry, which arises purely from $B$-field contributions.

\end{abstract}

\end{titlepage}

\newcommand{\be}{\begin{equation}}
\newcommand{\ee}{\end{equation}}
\newcommand{\bes}{\begin{equation*}}
\newcommand{\ees}{\end{equation*}}
\newcommand{\bea}{\begin{eqnarray}}
\newcommand{\eea}{\end{eqnarray}}
\newcommand{\beas}{\begin{eqnarray*}}
\newcommand{\eeas}{\end{eqnarray*}}
\newcommand{\bb}{\mathbb}
\newcommand{\tr}{\textrm{Tr}}
\newcommand{\del}{\nabla}
\newcommand{\myatop}[2]{\genfrac{}{}{0pt}{}{#1}{#2}}
\newcommand{\ph}{\phantom}
\newcommand{\mc}{\mathcal}
\newcommand{\bp}{\bar{\partial}}
\newcommand{\p}{\partial}
\newcommand{\ti}{\tilde}
\newcommand{\ep}{\epsilon}
\newcommand{\bep}{\bar{\epsilon}}
\newcommand{\one}{\mathbf{1}}
\renewcommand{\tfrac}[2]{{\textstyle\frac{#1}{#2}}}
\newcommand{\sbinom}[2]{{\textstyle\binom{#1}{#2}}}
\newcommand{\tsum}{{\textstyle\sum}}
\newcommand{\msum}[2]{{\displaystyle\sum_{#1}^{#2}}}
\newcommand{\tth}{\tilde{\theta}}
\renewcommand{\a}{\alpha}
\newcommand{\rt}{{\sqrt 2}}
\newcommand{\bbar}{\overline}
\newcommand{\ads}{{$AdS_3$}}
\newcommand{\wads}{{$\mathit{WAdS}_3$}}
\newcommand{\SL}{SL(2,\mathbb{R})}

\newcommand{\g}{{\gamma}}
\newcommand{\gb}{{\bar{\gamma}}}
\newcommand{\Nb}{{\overline{N}}}
\newcommand{\mb}{{\overline{m}}}
\newcommand{\U}{{\cal U}}
\newcommand{\V}{{\cal V}}

%%%%%%%%%%%%%%% Definitions Stephane%%%%%%%%%%%%%%%%%%%%%
\def\Q#1#2{\frac{\partial #1}{\partial #2}}
\def\varQ#1#2{\frac{\delta #1}{\delta #2}}
\def\norm#1{\|#1\|}
\def\eps{\epsilon}
\def\vareps{\varepsilon}
\def\d{\partial}
\def\cA{{\cal A}}
\def\cB{{\cal B}}
\def\cC{{\cal C}}
\def\cD{{\cal D}}
\def\cE{{\cal E}}
\def\cF{{\cal F}}
\def\cG{{\cal G}}
\def\cH{{\cal H}}
\def\cI{{\cal I}}
\def\cJ{{\cal J}}
\def\cK{{\cal K}}
\def\cL{{\cal L}}
\def\cM{{\cal M}}
\def\cN{{\cal N}}
\def\cO{{\cal O}}
\def\cP{{\cal P}}
\def\cQ{{\cal Q}}
\def\cR{{\cal R}}
\def\cS{{\cal S}}
\def\cT{{\cal T}}
\def\cU{{\cal U}}
\def\cV{{\cal V}}
\def\cW{{\cal W}}
\def\cX{{\cal X}}
\def\cY{{\cal Y}}
\def\cZ{{\cal Z}}

\def \ra{{\rightarrow}}
\def \lra{{\longrightarrow}}
\def \Ra{{\Rightarrow}}
\def \tr{{transformations\,\,}}
\def \CT{{conformal transformations\,\,}}
\def \eqdef{{\,\, \overset{\triangle}{=} \,\,}}
\def \gmu{{g_{\mu\nu}}}
\def \p{{\partial}}
\def \pb{{\overline{\partial}}}
\def \var{{\varepsilon}}
\def \eps{{\epsilon}}
\def \zb{{\overline{z}}}
\def \wb{{\overline{w}}}
\def \C{{\mathbb C}}
\def \R{{\mathbb R}}
\def \Z{{\mathbb Z}}
\def \abar{{\bar{\alpha}}}
\newcommand{\re}[1]{(\ref{#1})}
\newcommand{\bra}[1]{\langle \, #1 \, |}
\newcommand{\ket}[1]{|\, #1 \, \rangle}
\newcommand{\ke}[1]{|\, #1 \, \rangle}
\def\hb{{\bar{h}}}
\def\cd{{\cal D}}
\def\ex{{\mathrm{e}}}
\def\cL{{\cal L}}

%%%%%%%%%%%%%%%%%%%%%%%%%%%%%%%%%%%%%%%%%%%%%%%%%

\section{Introduction and Summary}

There are many reasons to believe that holography applies more broadly than just to anti-de Sitter spacetimes.  Indeed, recently holography has related certain non-relativistic CFT$_d$ to Schrodinger backgrounds in $d+1$ dimensions
\cite{Guica:2010sw, Son:2008ye, Balasubramanian:2008dm}, and spacelike warped $AdS_3$ has made an appearance in the near horizon geometry of extremal Kerr to which a dual CFT was conjectured \cite{Guica:2008mu}.  As a relatively simple extension of $AdS_3$, warped $AdS_3$ spacetimes appear to provide a useful playground for testing the bounds of holography \cite{Anninos:2008fx,Detournay:2010rh}, but an extension of the Brown-Henneaux analysis of asymptotic symmetries in $AdS_3$ is challenging since the formalism follows a `guess-and-check' method for choosing interesting boundary conditions \cite{Brown:1986nw}.

In an attempt to shed light on `interesting' boundary conditions, the present authors considered an embedding of warped $AdS_3$ into an exact string background \cite{Detournay:2010rh}, where the full power of a worldsheet CFT could be brought to bear.  What we found was rather surprising: for irrational values of the warping parameter (in appropriate units), the spacetime symmetries consist of a single Virasoro algebra, $\bar{L}^{\mathrm{st}}_m$, and a commuting global $U(1)$ charge $L^{\mathrm{st}}_0$; for rational values of the warping parameter, the full two-dimensional conformal symmetry is restored.

The holographic interpretation of this phenomenon is unclear at present.  The difficulty, of course, is that the warping of $AdS_3$ changes the asymptotics of the spacetime, corresponding to a deformation of the dual theory by an irrelevant operator which therefore changes the ultraviolet fixed point, an operation that we seemingly cannot understand via RG (though for the case of null-warped $AdS_3$, it was understood through appeal to certain additional symmetries \cite{Guica:2010sw}). In a similar setup, the authors of \cite{ElShowk:2011cm,Song:2011sr} used the fact that these backgrounds can be obtained as a TsT ({\it s} stands for ``shift'') transformation acting on the undeformed background to suggest that the dual theory is a dipole deformation of the original CFT, in analogy to higher-dimensional constructions such as \cite{Imeroni:2008cr}. What symmetries are preserved by the deformation remains unclear.

Here we opt to consider a deformation of the near horizon F1-NS5 system that is morally similar to that of our previous work, but without changing the asymptotics of the noncompact directions of spacetime.  Beginning with an $\SL \times SU(2) \times U(1) \times U(1)^3$ WZW model (describing string propagation on $AdS_3 \times S^3 \times S^1 \times T^3$), we modify the action by an exactly marginal deformation of the form \cite{Forste:2003km}
\begin{equation}
\Delta S \propto \textsc{h} \int d^2 z \, K^3(z) \bar{\partial}\varphi(\bar{z}) \, ,
\end{equation}
where $K^3(z)$ is a holomorphic current from the $SU(2)$ sector of the theory and $\varphi$ is a coordinate on the $S^1$.  This has the effect of warping the $S^3$ while simultaneously fibering the $S^1$ over the newly-warped $S^3$.  Calling the total space of that bundle $X_4$, our spacetime becomes $AdS_3 \times X_4 \times T^3$  (This is the same procedure we followed in our earlier paper, but there $K^3$ was replaced by $J^3$, a current from the $\SL$ sector). Deformations of the NS5-F1 system from an exact worldsheet CFT point of view have been addressed in the past \cite{MariosPetropoulos:2005wu,Fotopoulos:2007rm,Petropoulos:2009jz}, and dealt with a slightly different kind of marginal operators. A subset of these worldsheet conformal field theory deformations admits an interpretation either as a geometric deformation of the NS5-brane system or as a deformation of the distribution of the F1-branes, viewed as smooth instantons, inside the wrapped NS5-brane worldvolume. 

Following the work of \cite{Giveon:1998ns,Kutasov:1999xu,Giveon:2001up}, we can identify operators in the worldsheet CFT that correspond to the spacetime asymptotic symmetry generators.  Since this deformation does not involve the $\SL$ sector of the theory, it does not affect the asymptotics of the noncompact directions of spacetime; we therefore have two commuting Virasoro algebras regardless of the value of $\textsc{h}$.  On the other hand, the deformation does affect the $SU(2)$ sector of the theory, generically breaking the affine $\widehat{SU(2)}\times\widehat{SU(2)}$ of the dual (spacetime) CFT to $\widehat{U(1)}\times\widehat{SU(2)}$.  Since the affine $\widehat{SU(2)}{}^2$ plays the role of the R-symmetry of the dual $(4,4)$ SCFT (see \cite{Seiberg:1999xz} and \cite{David:2002wn} for a review), this implies that the dual supersymmetry is also generically broken --- in fact, it's broken to $(0,4)$ even though the R-symmetry analysis could have allowed a $(2,4)$ superconformal algebra.  However, as in our study of warped $AdS_3$, we find a restoration of the dual supersymmetry to $(4,4)$ whenever the warping parameter $\textsc{h}$ is integral in appropriate units.  From the ten-dimensional viewpoint, the meaning of this enhancement is perfectly clear as one can see that it is precisely at these points that $X_4$ can be written globally as a product of $S^3$ and a circle (with a different radius than the original $S^1$). On the other hand, if one were to dimensionally reduce and truncate to a three-dimensional theory on $AdS_3$, the enhancement would be unexpected since the $SU(2)$ gauge fields of the undeformed theory become massive when $\textsc{h}\neq 0$.  What is happening is the same as in the usual story of string compactifications on a circle: one of the higher Kaluza-Klein modes of each of the gauge fields, which were massive when $\textsc{h}=0$, become massless at each enhancement points.  We explain this in detail is section \ref{sec:gaugef}. From the viewpoint of the dual CFT, we are turning on a deformation of the theory that breaks $(4,4)$ supersymmetry to $(0,4)$, but (re-)enhances to $(4,4)$ at special values.  We demonstrate exactly how this enhancement occurs in the singular D1-D5 SCFT, finding that the deformation can be equivalently realized as a mixing between the $\mathbb{Z}$-orbifold defining one of the $T^4$ coordinates with phase rotations of the fermions.  The special values of the deformation parameter correspond to those points where the fermions (and hence the supercurrents) are invariant under the action of this $\mathbb{Z}$-orbifold.

In section \ref{sec:sugra}, we setup the supergravity background and check the dependence of supersymmetry on the deformation parameter.  In section \ref{sec:WZW}, we use worldsheet techniques to find the dependence of the spacetime $\widehat{SU(2)}$ R-symmetry on the deformation parameter, commenting on the construction of spacetime supercharges from worldsheet operators.  In section \ref{sec:gaugef}, we discuss how a naive dimensional-reduction and truncation to three-dimensions is blind to the supersymmetry enhancement and explain how to properly view the three-dimensional picture.  In section \ref{sec:dualCFT}, we describe the deformation of the singular D1-D5 CFT that corresponds to warping $S^3$ and show how supersymmetry is restored at quantized values of the deformation parameter.  In section \ref{sec:summary}, we summarize our results and discuss future directions.  Finally, it's worth mentioning that in appendix \ref{app:affineSU2}, we perform an asymptotic symmetry analysis to properly calculate the level of the asymptotic affine $\widehat{SU(2)}$ symmetry.  As is required by supersymmetry, the level is $\frac{c}{6} = \frac{\ell}{4G_3}$; interestingly, unlike the Brown-Henneaux calculation and the other calculations of central terms with which we are familiar, the level arises entirely from $B$-field contributions rather than metric contributions.

%%%%%%%%% SETUP AND SUGRA %%%%%%%%%%%%%%

\section{Setup and SUGRA Analysis}
\label{sec:sugra}

A fundamental string living in the near-horizon geometry of the F1-NS5 system (geometrically $AdS_3 \times S^3 \times T^4$) is described by an $\SL \times SU(2) \times U(1)^4$ WZW model \cite{Raju:2007uj, Pakman:2009mi, Giribet:2007wp, Dabholkar:2007ey, Pakman:2007hn}. The deformation we'll study acts only on an $SU(2)\times U(1)$ factor of the WZW model:
\begin{equation}
\Delta S \propto \textsc{h} \int d^2 z \, K^3(z) \bar{\partial}\varphi(\bar{z}) \, ,
\end{equation}
where $K^3(z)$ is a holomorphic current from the $SU(2)$ sector of the theory and $\bar{\partial}\varphi(\bar{z})$ is an antiholomorphic current from the $U(1)$ sector ($\varphi$ is a coordinate on an $S^1$ with radius $R$).  This has the effect of warping the $S^3$, which we denote by $\widetilde{S}^3$, while simultaneously fibering the $S^1$ over the $\widetilde{S}^3$.  Calling the total space of this bundle $X_4$ ($S^1\hookrightarrow X_4 \stackrel{\pi}{\longrightarrow} \widetilde{S}^3$), our spacetime becomes $AdS_3 \times X_4 \times T^3$.  The metric of $X_4$ can be written as
\begin{eqnarray}\label{x4metric}
ds^{2}_{X_{4}}=\frac{k}{4}\left[d\beta^{2}+\sin^{2}\beta d\alpha^{2}+(1-\textsc{h}^{2})(d\gamma +\cos \beta d\alpha)^{2}\right]+\left[\frac{\sqrt{k}\ \textsc{h}}{2}(d\gamma +\cos \beta d\alpha)+d\varphi\right]^{2} \, ,
\end{eqnarray}
where $\textsc{h}\in[0,1)$.
The undeformed WZW model already had a $B$-field whose flux, $H_3$, was a linear combination of the $AdS_3$ and $S^3$ volume forms.  The deformation modifies this to
\begin{eqnarray}
B=B_{AdS_{3}}+\frac{k}{4}\cos\beta d\alpha\wedge d\gamma+\sigma\frac{\sqrt{k}\textsc{h}}{2}(d\gamma +\cos \beta d\alpha)\wedge d\varphi  \, ,  \nonumber\\
H_{3}=H_{AdS_{3}}-\frac{k}{4}\sin\beta d\beta\wedge d\alpha\wedge d\gamma-\sigma\frac{\sqrt{k}\textsc{h}}{2}\sin \beta d \beta \wedge d\alpha\wedge d\varphi  \, ,
\end{eqnarray}
where $\sigma=1$ (if we instead deformed by $\overline{K}{}^{3}\partial \varphi$, then $\sigma$ would be $-1$, an important distinction particularly for the heterotic theory where the sign of $\sigma$ determines how much spacetime supersymmetry will be broken by the deformation).

The simplicity of this deformation is that the effect on the worldsheet CFT can be entirely understood as a rotation on the toroidal lattice defined by $\varphi$ and by the scalar one obtains from bosonizing $K^3$ \cite{Forste:2003km} (essentially the coordinate on the $S^1 \subset S^3$ of the Hopf fibration).  This is reflected in the fact that the metric (\ref{x4metric}) is {\it locally} isomorphic to $S^{1}\times S^{3}$. This can be seen by defining the new variables
\begin{eqnarray}
\label{eqn:coordxform}
\varphi'=\sqrt{1-\textsc{h}^{2}}\varphi \, , \qquad \gamma'=\gamma+\frac{2\textsc{h}}{\sqrt{k}}\varphi \, ,
\end{eqnarray}
where the identifications $\gamma \cong \gamma + 2\pi$ and $\varphi \cong \varphi + 2\pi R$ become
\begin{eqnarray}\label{identx4}
\gamma'\cong\gamma'+2\pi \, , \qquad
(\gamma',\varphi')\cong (\gamma',\varphi') + 2\pi R \left(\frac{2\textsc{h}}{\sqrt{k}}\ , \, \sqrt{1-\textsc{h}^{2}}\right) \, .
\end{eqnarray}
In particular, notice that whenever $\frac{2\textsc{h}R}{\sqrt{k}}\in\mathbb{Z}$ the second identification only acts nontrivially on $\varphi'$ so that $X_4$ is {\it globally} isomorphic to $S^3 \times S^1$, but the radius of this $S^1$ is $R\sqrt{1-\textsc{h}^2}$, different from the fibered $S^1$.
Another special case occurs when $\frac{2\textsc{h}R}{\sqrt{k}}\in \mathbb{Q}$; calling $\frac{2\textsc{h}R}{\sqrt{k}} \equiv \frac{\mu}{\rho}$, we see that the theory is a $\mathbb{Z}_\rho$-orbifold (with discrete torsion from a total derivative term) of an undeformed theory with $S^1$ radius $\mu R \sqrt{1-\textsc{h}^2}$.  The action of $\mathbb{Z}_\rho$ extends to the full F1-NS5 system and so in these cases the dual can be understood as a $\mathbb{Z}_\rho$-orbifold of an undeformed theory with a different $S^1$ radius.

One can also analyze the supersymmetry of the background, which is virtually the same as the analysis in \cite{Detournay:2010rh}.  The vanishing of the $\varphi$ component of the IIA gravitino variation implies
\begin{eqnarray}
\left(\partial_{\varphi}+\frac{\textsc{h}}{\sqrt{k}}(1\pm \sigma)\Gamma^{\hat{1}\hat{2}}\right)\varepsilon_{\pm} = 0 \, ,
\end{eqnarray}
where the vielbein directions are $e^{\hat{1}}=\frac{\sqrt{k}}{2}d\beta$ and   $e^{\hat{2}}=\frac{\sqrt{k}}{2}\sin \beta d\alpha$. One of the Killing spinors, $\varepsilon_{\sigma}$, will then only be globally well-defined (invariant under shifts of $\varphi$) when
\begin{eqnarray}
\label{eqn:resonance}
\frac{2\textsc{h}R}{\sqrt{k}}\in \mathbb{Z} \, ,
\end{eqnarray}
otherwise half of supersymmetry is broken for IIA so that the $(4,4)$ supersymmetry of the dual SCFT will be broken to $(0,4)$ supersymmetry.  In the heterotic case, we only have one of the $\varepsilon_{\pm}$ spinors to begin with, so depending on the sign of $\sigma$, either no supersymmetry is broken or all of supersymmetry is broken when (\ref{eqn:resonance}) fails to hold.  Type IIB has the same qualitative results as IIA.  Not surprisingly, we see that the condition for unbroken supersymmetry is equivalent to the condition that $X_4$ be globally isomorphic to $S^3 \times S^1$.

%%%%%%%%%%%%%%%% DEFORMED WZW MODEL %%%%%%%%%%%%%%

\section{Deformed WZW Model and Spacetime $\widehat{SU(2)}$}\label{sec:defWZW}
\label{sec:WZW}

The situation in this case is simpler than for warped $AdS_{3}$, particularly because the spectral flow in $SU(2)$ WZW theories is an inner automorphism, so it won't mix representations (for example, see \cite{Giribet:2007wp}).  Following \cite{Detournay:2010rh}, deforming the theory by a spacelike current results in a Lorentzian rotation of the charge lattice defining primary operators.  We find the analysis simplest by using the BRST formalism for the coset CFT \cite{Karabali:1989dk,Hwang:1993nc}; in order to keep the analysis more clear, we will focus on the bosonic case, but the results carry over simply to the supersymmetric case, as they did in \cite{Detournay:2010rh}.

We can express an $SU(2)_k$ WZW model as a discrete quotient of $\big(SU(2)_k / U(1) \big) \times U(1)$, where the $U(1)$ in the numerator is generated by the $K^3$ that we use in the deformation.  We then realize the coset $SU(2)_{k}/U(1)$ as $\big(SU(2)_k \times U(1) \times (bc) \big)/\sim$, where $(bc)$ is a ghost system with weights 1 and 0 (not to be confused with the ghosts necessary for gauge-fixing worldsheet gravity) and $\sim$ is defined through a BRST operator $Q_{G/H}$ (again, this BRST operator is in addition to the BRST operator defining the physical states in the full string theory).  The $\widehat{U(1)}$ of the BRST formulation, generated by $\widehat{K}_H$,  must have ``level'' $-\frac{k}{2}$, opposite that of $K^3$. We bosonize the currents $K^3 = i\sqrt{\frac{k}{2}}\, \partial Y$ and $\widehat{K}_H = i\sqrt{\frac{k}{2}}\,\partial T$, with $Y$ a spacelike boson and $T$ a timelike boson.
The BRST operator for the coset theory can then be expressed as
\begin{eqnarray}
Q_{G/H} \equiv \oint \frac{d z}{2\pi i} :\! c \big( K^{3} + \widehat{K}_{H} \big) \! : \  \, .
\end{eqnarray}
A primary state in the BRST formulation is annihilated by $c_{n>0}, \, b_{n\geq 0}, \, K^a_{n>0},$ and $\widehat{K}_{H,n>0}$.  Thus, a \emph{physical} primary state without ghost excitations must have equal and opposite eigenvalues of $K^3_{0}$ and $\widehat{K}_{H,0}$.

First, let's take a brief look at the undeformed case. In order to have $(1,0)$ holomorphic operators which can be used to construct spacetime currents we must set $j=0=\bar{m}=\overline{N}$ (where $\frac{j}{2}$ is the ``spin'' of the representation, $m$ and $\bar{m}$ are the $J^3$ and $\bar{J}^3$ eigenvalues, and $N$ and $\overline{N}$ are the number of oscillators excited). $SU(2)$ representations with $j=0$ have $m=0$, so the most general operator we can construct from the $SU(2)\times U(1)$ sector is
\begin{eqnarray}\label{vertexundef}
 \left[ a\partial Y+b^{+}K^{+}e^{-i\sqrt{\frac{2}{k}}(T+Y)}+b^{-}K^{-}e^{i\sqrt{\frac{2}{k}}(T+Y)} +c\partial T + d\partial\varphi \right] \, ,
\end{eqnarray}
where $\varphi$ generates the additional $\widehat{U(1)}$ and has radius $R$ (we will only consider generic $R$, where we need not worry about enhancements such as at the self-dual radius).  These currents can then be lifted to affine spacetime currents if we multiply them by an appropriate operator from the $\SL$ sector of the theory, given by $\mathcal{V}_{\widetilde{m}}e^{-i\widetilde{m}\sqrt{\frac{2}{k}}(\phi_{t}+\phi_{s})}$ (for their exact form, see the Appendix D of \cite{Detournay:2010rh}).

Deforming our theory by adding to the Lagrangian an operator proportional to $K^{3}\overline{\partial} \varphi$ will change the conformal dimension of (\ref{vertexundef}), so we modify it by multiplying by $e^{ip_L\varphi_L + ip_R\varphi_R}$.  The deformation induces a rotation on the charge lattice
\be
\left(\begin{array}{c} m \\ p_R \end{array}\right) \longmapsto \left(\begin{array}{c} m' \\ p'_R \end{array}\right) = \left(\begin{array}{c} m\cosh\alpha+\sqrt{\frac{k}{2}}p_{R}\sinh\alpha  \\  p_{R}\cosh\alpha+\sqrt{\frac{2}{k}}m\sinh\alpha \end{array}\right) \ee
where $p_{L,R} = \frac{p}{R} \pm \frac{wR}{2}$, $p$ corresponds to $S^1$ momentum and $w$ to $S^1$ winding, and the deformation parameter $\textsc{h}$ is related to $\alpha$ by
\begin{equation}
\cosh^{2}\alpha=\frac{1}{1-2\textsc{h}^{2}} \, .
\end{equation}
The operators of the deformed theory include terms of the form $e^{i\sqrt{\frac{2}{k}}\left(m' \widehat{Y}_L + \overline{m}'\widehat{Y}_R\right)+ip_L'\widehat{\varphi}_L + ip'_R\widehat{\varphi}_R}$, where $\widehat{Y}$ and $\widehat{\varphi}$ have canonical OPEs.  For holomorphy, then, we should set $\overline{m}' = \overline{m}=0$ and $p'_{R}=0$, or
\begin{equation}
p_{R}=-\sqrt{\frac{2}{k}}m\tanh\alpha  \, .
\end{equation}
Restricting to the sector with no winding, $w=0$, then $p_L=p_R$ and these vertex operators become
\begin{equation}
\label{operator}
e^{i\sqrt{\frac{2}{k}}m\,\mathrm{sech}\,\alpha\,\widehat{Y}_L-i\sqrt{\frac{2}{k}}m\tanh\alpha\, \widehat{\varphi}_L} \, .
\end{equation}
We can simplify further by defining a new set of operators $(\widetilde{Y},\widetilde{\varphi})$ as
\begin{equation}
\label{rotationbosons}
\widetilde{Y}=\mathrm{sech}\,\alpha\,\widehat{Y}-\tanh\alpha\,\widehat{\varphi} \, ,  \qquad  \widetilde{\varphi}=\mathrm{sech}\,\alpha\,\widehat{\varphi}+\tanh\alpha\, \widehat{Y} \, ,
\end{equation}
so the operators (\ref{operator}) reduce to
\begin{equation}
\label{operator2}
e^{i\sqrt{\frac{2}{k}}m\widetilde{Y}_L}\, .
\end{equation}
For the purposes of identifications, we can schematically relate the new variables to the old ones as
\begin{eqnarray}
\widetilde{Y}_L = Y_L - \big(\varphi_L + \varphi_R\big)\tanh\alpha \, ,  &  \qquad  &  \widetilde{\varphi}_L = \varphi_L \,\mathrm{sech}\,\alpha - \varphi_R \sinh\alpha\tanh\alpha + Y_L \sinh\alpha \, ,  \nonumber \\
\widetilde{Y}_R = Y_R \cosh\alpha + Y_L \sinh\alpha\tanh\alpha - \varphi_R \sinh\alpha \, ,  &  \qquad &  \widetilde{\varphi}_R = \varphi_R + \big(Y_R - Y_L\big) \tanh\alpha \, .
\end{eqnarray}
Thus, the original identifications become
\begin{eqnarray}
\label{eqn:identifications}
\big(\widetilde{Y}_L , \widetilde{Y}_R ; \widetilde{\varphi}_L, \widetilde{\varphi}_R \big) &\cong &  \big(\widetilde{Y}_L , \widetilde{Y}_R ; \widetilde{\varphi}_L, \widetilde{\varphi}_R \big) + 2\pi\sqrt{\tfrac{k}{2}} \, \big(  1 , 2\cosh\alpha-\mathrm{sech}\,\alpha ; \, \sinh\alpha , 0 \big) \nonumber \\
&\cong & \big(\widetilde{Y}_L , \widetilde{Y}_R ; \widetilde{\varphi}_L, \widetilde{\varphi}_R \big)   +   \pi R \big( -2\tanh\alpha, -\sinh\alpha ; \, \mathrm{sech}\,\alpha , 1 \big)  \, .
\end{eqnarray}
Since $\tanh\alpha = \sqrt{2}\,\textsc{h}$, the operators (\ref{operator2}) will be well defined for any $m\in\mathbb{Z}$ under the second identification when
\begin{eqnarray}
\frac{2\textsc{h}R}{\sqrt{k}}\in \mathbb{Z} \, ,
\end{eqnarray}
which is the same condition obtained before from the supergravity analysis.  When this condition holds, we can build generators of a worldsheet $\widehat{SU(2)}_k$ out of $\p\widetilde{Y}$, $K^+ e^{-i\sqrt{\frac{2}{k}}(T+\widetilde{Y}_L)}$, and $K^- e^{i\sqrt{\frac{2}{k}}(T+\widetilde{Y}_L)}$, all of which survive the identifications (\ref{eqn:identifications}).
%When this condition holds, we can build generators of a worldsheet $\widehat{SU(2)}_k$ out of $\p\widetilde{Y}$, $e^{\pm i \sqrt{\frac{2}{k}}\,\widetilde{Y}_L}$, and the parafermions of $\widehat{SU(2)}_k/\widehat{U(1)}$ since they are unaffected by the deformation of the worldsheet theory.
When $\frac{2\textsc{h}R}{\sqrt{k}} \equiv \frac{\mu}{\rho}\in\mathbb{Q}$, the operators (\ref{operator2}) will be well-defined for $m\in\rho\mathbb{Z}$, however there is no operator in $\widehat{SU(2)}_k/\widehat{U(1)}$ with the correct conformal dimension that can be paired with (\ref{operator2}) when $\rho\neq 1$ (indeed, for $m^2>k$ this operator would need to have negative conformal dimension).  Therefore, for a generic deformation we cannot construct spacetime charges from the operators (\ref{vertexundef}) unless $b^{\pm}=0$, then instead of having two copies of $\widehat{SU(2)}$ in the spacetime SCFT we are left with only $\widehat{U(1)}\times\widehat{SU(2)}$.  Since the $SU(2)$ symmetry acts as the R-symmetry of the dual theory, we expect some of the spacetime supersymmetries to be broken as well, in agreement with section \ref{sec:sugra}.

The same computation as in \cite{Giveon:1998ns} (explained more throughly in \cite{deBoer:1998pp} and \cite{Kutasov:1999xu}) demonstrates that the level of the spacetime $\widehat{SU(2)}$ receives a contribution of $k$ for each time the worldsheet wraps the spatial $S^1$ of $AdS_3$ --- the interpretation is that the worldsheet acts as a domain wall in the $AdS_3$ background, with discontinuity in the level of the spacetime $\widehat{SU(2)}$ given by $k\times(\textit{winding})$.  The contribution to the level of the spacetime $\widehat{SU(2)}$ from this discontinuity is separate from the contribution made by the original background; the discontinuity arises from long strings probing the background, while the original background creates contributions visible to local metric fluctuations or, equivalently, to short strings.  A supergravity asymptotic symmetry analysis (in the vein of Brown and Henneaux) is a study of metric fluctuations; a computation of the equivalent affect from the viewpoint of short strings was begun in \cite{deBoer:1998pp,Kutasov:1999xu}, but only completely very recently in \cite{Troost:2011ud} for the Virasoro central charge.  In appendix \ref{app:affineSU2}, we perform an asymptotic symmetry analysis on the supergravity background and obtain the result expected from the $(4,4)$ supersymmetry of the dual, namely $k_{\mathrm{st}} = \frac{c_{\mathrm{st}}}{6} = \frac{\ell}{4G_3}$.  The novel feature of the analysis is that the value for the level comes entirely from contributions of the $B$-field, unlike typical computations for Virasoro central terms which only receive contributions from the metric.

The construction of spacetime supersymmetries from worldsheet operators parallels the discussion in \cite{Detournay:2010rh}. The main difference is that in the present case we will define the rotated fermions analogously to (\ref{rotationbosons}), winding up with a pair of rotated fermions $\widetilde{\Psi}^{3}$ and $\widetilde{\Psi}^{Z}$.  We could either choose to bosonize the pair $(\widetilde{\Psi}^{3},\widetilde{\Psi}^{\varphi})$, or we could bosonize $\widetilde{\Psi}^{3}$ with some other fermion (e.g., one of the free fermions of the $\SL$ WZW model), the conclusion will be the same: the existence of spacetime supercharges constructed from holomorphic worldsheet operators is only possible if $\frac{2\textsc{h}R}{\sqrt{k}}\in \mathbb{Z}$. Therefore, any generic deformation will break half of the supersymmetries in type II theories, in agreement with our supergravity analysis.

\section{Spacetime Gauge Fields from WS Currents}
\label{sec:gaugef}

We have seen from the worldsheet and from ten-dimensional supergravity that there is an enhancement of the symmetry of the dual theory whenever $\frac{2\textsc{h}R}{\sqrt{k}}\in\mathbb{Z}$.  In particular, there is an enhancement of a $U(1)$ factor to an $SU(2)$.  Holography tells us that from the perspective of the dimensionally-reduced three-dimensional theory on $AdS_3$, there must be massless $SU(2)$ gauge fields at these enhancement points, and that two of the gauge fields must become massive as we move away from the points of enhancement.  We will demonstrate that this is the case and compute the masses using worldsheet techniques.

It is well known that whenever we have holomorphic $(1,0)$ currents in a worldsheet theory, these give rise to gauge fields in spacetime.  In a CFT that is $\mathbb{R}^d \times \mathcal{M}_{\mathrm{int}}$, a $(1,0)$ holomorphic current $K(\sigma)$ in the internal CFT leads to a target-space gauge field
\begin{equation}
\mathbf{A}_\mu[X(\sigma)] \propto \int d^d\!k\, d^2\!\sigma\,  e^{i k\cdot X(\sigma)} \tilde{A}_\mu(k) K(\sigma) \bar{\partial} X^\mu(\sigma) \, \delta(k^2) \, ,
\end{equation}
where the delta-function in $k^2$ is necessary to ensure that the operator is marginal.
This vertex operator thus corresponds to a massless vector field.
If we deform the internal CFT by an exactly marginal deformation, the dimension of $K$ may change to, say, $(1+\gamma,\gamma)$,
in which case the vertex operator must be modified to
\begin{equation}
\mathbf{A}_\mu[X(\sigma)] \propto \int d^d\!k\, d^2\!\sigma\,  e^{i k\cdot X(\sigma)} \tilde{A}_\mu(k) K(\sigma) \bar{\partial} X^\mu(\sigma) \, \delta(k^2+2\gamma)
\end{equation}
in order to be marginal, so the gauge field acquires a mass
\begin{equation}
\mathfrak{m}^2 \propto 2\gamma \, .
\end{equation}

\subsection{Gauge Fields in $AdS_3$}

Now we'd like to replace $\mathbb{R}^d$ by $AdS_3$.  For this, we use the results and notation from appendix \ref{app:reps}.  The isometry group of $AdS_3$ is composed of the Lorentz generators $\mathbf{L}$ as well as would-be translations $\mathbf{R}$.  $\mathbf{R}^2$ is the Laplacian $\frac{1}{\Lambda}\nabla^2$, so its eigenvalue $\mathfrak{r}$ is directly related to $\frac{\mathfrak{m}^2}{\Lambda}$.  To get the precise relationship, consider the equation of motion for a massive vector field
\begin{equation}
\nabla^2 A_\mu - \nabla^\nu \nabla_\mu A_\nu - \mathfrak{m}^2 A_\mu = 0 \, ,
\end{equation}
which implies that $\mathfrak{m}^2 \nabla^\mu A_\mu = 0$.  If $\mathfrak{m}=0$, this condition tells us nothing, but we then have a gauge invariance that we can partially fix by choosing $\nabla^\mu A_\mu=0$, so for any $\mathfrak{m}$ we can simplify this equation to
\begin{equation}
\nabla^2 A_\mu - (2\Lambda + \mathfrak{m}^2) A_\mu = 0 \, ,
\end{equation}
where $\Lambda$ is the cosmological constant. Thus, we find that $\mathfrak{r} = 2 + \frac{\mathfrak{m}^2}{\Lambda}$.

We can now discuss gauge fields in $AdS_3$.  We would like to write down
\begin{equation}
\mathbf{A}[X(\sigma)] \sim \sum_{j,\bar{j},m,\bar{m},N,\overline{N}} c_{j\bar{j}m\bar{m}N\overline{N}}\int d^2\!\sigma\, K \,\mathcal{V}_{j\bar{j}m\bar{m}N\overline{N}} \ ,
\end{equation}
where $\mathcal{V}$ is a tensor of the $SL(2;\mathbb{R})$ WZW model with ``spins'' $\frac{j}{2}$ and $\frac{\bar{j}}{2}$, $J^3$ and $\bar{J}^3$ have eigenvalues $m$ and $\bar{m}$, and oscillator excitations are denoted by $N$ and $\overline{N}$.  For this to be a vector of the Lorentz group, we must set $-4j(j-1) = \mathfrak{r}-2$ and $m+\bar{m}=\hat{\mu} \in \{-1,0,1\}$ (see appendix \ref{app:reps}).  This operator should be marginal to correspond to a physical excitation of the theory, so we have
\begin{eqnarray}
\Delta_L &=& 1-\frac{j(j-1)}{k-2} + N = \frac{\mathfrak{r}-2}{4(k-2)} + N = 1  \, ,  \\
\Delta_R &=& \frac{\mathfrak{r}-2}{4(k-2)} + \overline{N} = 1 \, .
\end{eqnarray}
We should evidently set $\mathfrak{r}=2$, $N=0$, and $\overline{N}=1$.  Considering the above identification of $\mathfrak{r} = 2+\frac{\mathfrak{m}^2}{\Lambda}$, we see that this indeed corresponds to a massless gauge field:
\begin{equation}
\label{eqn:AAdS}
\mathbf{A}_{\hat{\mu}}[X(\sigma)] = \sum_{m} c_{\hat{\mu}m}\int d^2\!\sigma\, K \,\mathcal{V}_{j=\bar{j}=1,m,\bar{m}=\hat{\mu}-m,N=0,\overline{N}=1} \, .
\end{equation}

The deformed case is now straightforward, but before we proceed let's promote our theory to a superconformal WZW model, which implies we must replace the factor of $(k-2)$ appearing in denominators by $k$.  For the case that the internal SCFT is a deformed $SU(2)\times U(1)$ WZW model, the $SU(2)$ currents $K^\pm$ have their dimensions deformed by\footnote{In the undeformed theory, we can write $J^{\pm} \propto \psi_{\pm}e^{\pm i\sqrt{2/k}Y_{L}}$, where $\psi_{\pm}$ is a parafermion of weight $1-\frac{1}{k}$ (see section \ref{sec:defWZW}).  After deforming, the holomorphic part of $Y$ transform as $Y_{L}\rightarrow Y_{L}\cosh \alpha-\varphi_{R}\sinh\alpha$, so the weights of the deformed $J^{\pm}$ become $\Delta_{L}(J^{\pm})=1+\frac{\sinh^2\alpha}{k}$ and $\Delta_{R}(J^{\pm})=\frac{\sinh^2\alpha}{k}$.}
\begin{equation}
\Delta(K^\pm): \quad (1,0)\longrightarrow \big( 1 + \tfrac{1}{k}\sinh^2\alpha , \, \tfrac{1}{k}\sinh^2\alpha \big) \, .
\end{equation}
Working in the NS-NS sector and $0$-picture, we can write the spacetime gauge field in the same way as (\ref{eqn:AAdS}), except that the condition of being marginal now imposes
\begin{eqnarray}
\Delta_L &=& 1+\frac{1}{k}\sinh^2\alpha + \frac{\mathfrak{r}-2}{4k} + N = 1  \, ,  \\
\Delta_R &=& \frac{1}{k}\sinh^2\alpha + \frac{\mathfrak{r}-2}{4k} + \overline{N} = 1 \, .
\end{eqnarray}
We should still choose $N=0$ and $\overline{N}=1$ so that this has the appropriate $\alpha\rightarrow 0$ limit.  We see then that
\begin{equation}
\mathfrak{r}-2 = \frac{\mathfrak{m}^2}{\Lambda} = -4\sinh^2\alpha \, ,
\end{equation}
so our gauge fields gain the mass $\mathfrak{m} = 2\sqrt{|\Lambda|} |\sinh\alpha| = \sqrt{|\Lambda|}\,\frac{2\sqrt{2}\textsc{h}}{1-2\textsc{h}^2}$ as we deform, never to be massless again.

To summarize, we have found that we have a set of spacetime $SU(2)$ gauge fields, $\mathbf{A}^3,\mathbf{A}^\pm$, that are massless when $\alpha=0$ but that reduce to a massless $U(1)$ gauge field $\mathbf{A}^3$ and two massive gauge fields $\mathbf{A}^\pm$ with $\mathfrak{m}=2\sqrt{|\Lambda|} |\sinh\alpha|$.  When $\alpha=0$, we know that these gauge fields are dual to the $SU(2)_R$ current of the dual CFT, but for all other $\alpha\neq 0$ they cannot be because the $\mathbf{A}^\pm$ are massive.  On the other hand, we know from both a worldsheet analysis and a ten-dimensional supergravity analysis that supersymmetry exists whenever $\frac{2\textsc{h}R}{\sqrt{k}}\in\mathbb{Z}$, so there should be massless $SU(2)$ gauge fields at these enhancement points.  From the earlier analysis, we know that when $\frac{2\textsc{H}R}{\sqrt{k}}\in\mathbb{Z}$, the operators $e^{\pm i\sqrt{\frac{2}{k}}\widetilde{Y}_L}$ are allowed by the identifications and we can use them to construct an affine $\widehat{SU(2)}$ algebra on the worldsheet at level $k$, giving rise to massless vectors in the spacetime theory.  As we move away from this point of enhanced symmetry, these gauge fields again become massive while other massive gauge fields are approaching zero mass as we move towards the next point of enhancement.  In looking more directly at spacetime supersymmetry, one could perform a similar analysis for the spacetime gravitino and would find the same result: the gravitino becomes massive as we move away from $\alpha=0$, but at enhanced points a previously-massive gravitino mode will become massless and restore supersymmetry.

This effect would be missed from a naive dimensional reduction and truncation to three-dimensions around the undeformed background.  What is happening is that the KK modes of the gravitino (and gauge fields) that we typically ignore in such a truncation have masses that vary continuously with our tuning of $\textsc{h}$.  Whenever $\frac{2\textsc{h}R}{\sqrt{k}}\in\mathbb{Z}$, one of the modes will be massless, but it will not be the mode that we would have kept when we expanded around $\textsc{h}=0$.  From the higher dimensional point of view, what is happening at these points is that the $S^1$ bundle over warped $S^3$ becomes expressible simply as a product $\hat{S}^1 \times S^3$, where $\hat{S}^1$ has a different radius than the $S^1$ of the fibration.  At these points, we could have instead chosen $\hat{S}^1$ for our KK-reduction/truncation, in which case we would have seen the supersymmetry restoration.  This is morally analogous to the enhancement seen when one tunes an $S^1$ to the self-dual point.

\section{Deforming the Dual SCFT}
\label{sec:dualCFT}

We have now understood the supersymmetry enhancement from both the worldsheet and spacetime perspectives, but understanding the enhancement from the dual CFT remains.  Evidently, we can turn on a marginal operator in the dual that breaks $\mathcal{N}=(4,4)$ supersymmetry to $\mathcal{N}=(0,4)$, but when the deformation parameter takes certain quantized values, $\mathcal{N}=(4,4)$ is restored.

To see this qualitative effect from the dual CFT, we will study the singular D1-D5 SCFT, which is given simply by a symmetric product of $Q_1 Q_5$ copies of the free $\mathcal{N}=(4,4)$ SCFT on $T^4$:  $(T^{4})^{Q_{1}Q_{5}}/S_{Q_{1}Q_{5}}$, where $Q_{a}$ is the number of D$a$-branes \cite{deBoer:1998ip,Dijkgraaf:1998gf,Seiberg:1999xz} (this is at a different point in the $\mathcal{N}=(4,4)$ moduli space than the dual of the D1-D5 background \cite{Dijkgraaf:1998gf}, but for our purposes it will be sufficient).  The undeformed action is
\begin{eqnarray}
\int d^{2}x\, \Big\{\partial X^{i}_{A}\bar{\partial}X^{i}_{A}+\psi^{i}_{A}(x)\bar{\partial}\psi^{i}_{A}(x)+\tilde{\psi}^{i}_{A}(\bar{x})\partial\tilde{\psi}^{i}_{A}(\bar{x})\Big\} \, ,
\end{eqnarray}
where $X^{i}_{A}$ are coordinates on the $A^{\textrm{th}}$ $T^{4}$ and $\psi$s are superpartners.  Fields can be organized according to their charges under $SU(2)_{R}\times SU(2)_{R}$ (from $S^3$ isometries in the bulk), $SO(4)$ (from the local $T^4$ isometries, even though they're broken by global identifications), and under the Virasoro algebra of the dual theory. Fortunately, we know what all of these quantum numbers are for the operator by which we deform: under $SU(2)_R \times SU(2)_R$ it transforms in the $(\mathbf{3},\mathbf{1})$ with $J^3_R$ eigenvalues $(0,0)$, under $SO(4)$ it transforms in the $\mathbf{4}$, and it must have conformal dimensions $(h_{\mathrm{st}},\bar{h}_{\mathrm{st}}) = (1,1)$.\footnote{These can be seen either directly from the worldsheet operators, or by acting on the deformations of the metric and $B$-field with the Lie derivatives corresponding to each isometry.}  A natural choice is then $\int d^2x \, J^3_R(x) \bar{J}(\bar{x})$, where $J^3_R$ is a holomorphic R-symmetry current and $\bar{J}$ is an antiholomorphic dimension $(0,1)$ current that transforms in the $(0,1)$ of $SO(4)$.\footnote{More generally, we could choose an operator of the form $\int d^2x\, \left(J^3_R(x)\right)_A \Omega^{AB} \left(\bar{J}(\bar{x})\right)_B$ where $(J)_A$ refers to the current $J$ on the $A^{\textrm{th}}$ $T^4$ and $\Omega$ commutes with $S_{Q_1 Q_5}$.  The supersymmetry enhancement will happen when the entries of $\Omega$ are rational multiples of each other, and we can always find an $\Omega$ satisfying these properties, as the choice made in the text does.  On the other hand, for choices of $\Omega$ other than the one in the main text, the (local) supercurrent will transform inhomogeneously under a shift along a single $T^4$, which seems unnatural to us.}  For some vector $V_i$, we can identify $\bar{J}$ with $\sum_A iV_i\bar{\partial} X^i_A$; for this qualitative analysis, it won't matter exactly what $V_i$ is.

Recalling that $J^{3}_{R}$ can be written in terms of free fields as \cite{David:2002wn}
\begin{eqnarray}
J^{3}_{R}=\frac{1}{2}\sum_A \big( \chi^{+}_{A}\bar{\chi}^{+}_{A}-\chi^{-}_{A}\bar{\chi}^{-}_{A}\big) \, ,
\end{eqnarray}
where $\chi^{+}_{A}\equiv\frac{1}{\sqrt{2}}(\psi^{1}_{A}+i\psi_{A}^{2})$ and $\chi^{-}_{A}\equiv\frac{1}{\sqrt{2}}(\psi^{3}_{A}+i\psi_{A}^{4})$, the action for the holomorphic fermions becomes
\begin{eqnarray}
2\sum_A \int d^{2}x\,\Big\{\chi^{+}_{A}\bar{\partial}\bar{\chi}^{+}_{A}+\chi^{-}_{A}\bar{\partial}\bar{\chi}^{-}_{A}  -  i\tilde{\textsc{h}}\big(\chi_A^+ \bar{\chi}_A^+ - \chi^-_A \bar{\chi}^-_A \big)\sum_{B}\big(V_i\bar{\partial}X^{i}_{B}\big)\Big\} \, ,
\end{eqnarray}
but we can remove the deformation by rotating the fermions
\be
\chi^{\pm}_{A}\ \longrightarrow \ \chi'^{\pm}_A \equiv e^{\pm i\tilde{\textsc{h}} V_i \sum_{B}X^{i}_{B}}\chi^{\pm}_{A} \, .
\ee
Being coordinates on $T^4$, we have four identifications $X^i_A \cong X^i_A + 2\pi R^i_{\ph{i}j}n^j$ where $\vec{n}\in\mathbb{Z}^4$, but we can always rotate by $SO(4)$ to choose coordinates where $R^1_{\ph{1}j} = R^1\delta^1_j$.  In these coordinates, we can choose $V_i = \delta^1_i$ since our goal is to show that their exists an operator in the dual with the correct quantum numbers that reproduces the qualitative features of supersymmetry enhancement.  Under this identification, when $X^1_B \rightarrow X^1_B + 2\pi R^1$ we see that 
\be
\label{eqn:dual-phase}
\chi'^{\pm}_A \rightarrow e^{\pm 2\pi i\tilde{\textsc{h}} R^1}\chi'^{\pm}_A
\ee
for all $A$.  With these free fields, we can write the complex supercurrents as $G^{+}\propto\sum_A \left(\chi'^{+}_{A}\partial Z^{-}_{A}-\bar{\chi}'^{-}_{A}\partial Z^{+}_{A}\right)$, $G^{-}\propto\sum_A\left(\chi'^{-}_{A}\partial Z^{-}_{A}+\bar{\chi}'^{+}_{A}\partial Z^{+}_{A}\right)$, as well as the other R-symmetry currents as $J^\pm_R = \sum_A \chi'^\pm_A \bar{\chi}'^\mp_A$, where $Z^\pm$ are defined similarly to $\chi^\pm$.  These operators will satisfy the $\mathcal{N}=4$ superconformal algebra, but we see immediately from (\ref{eqn:dual-phase}) that under a shift along $X^1$, $G^a$ and $J_R^\pm$ are not invariant and therefore are projected out of the spectrum of the theory, unless $\tilde{\textsc{h}}R^1\in\mathbb{Z}$ in which case they survive the projection and supersymmetry is restored.  This is the same qualitative result we obtained from the bulk viewpoint.

\section{Discussion}
\label{sec:summary}

Since the advent of the $AdS$/CFT correspondence, many steps have been taken towards describing more realistic field theories, for example by breaking supersymmetries or attempting to break conformal invariance.  One such example is the Lunin-Maldacena deformation of $AdS_5 \times S^5$ \cite{Lunin:2005jy}, which describes a marginal deformation of $\mathcal{N}=4$ SYM to an $\mathcal{N}=1$ theory. The dual gravity background is obtained as a TsT transformation that acts on two commuting isometries of the $S^5$, resulting in a squashed sphere whose $U(1)\times U(1)$ isometries realize the R-symmetry of the gauge theory geometrically.  Another set of applications that have emerged in recent years (with applications to Kerr/CFT or holographic condensed matter, for instance) involve non-$AdS$ spaces in a crucial way, but the dual theory in those cases is harder to identify and study (see however \cite{Guica:2010ej, Song:2011ii, Compere:2010uk} and \cite{Maldacena:2008wh,Guica:2010sw}). In the former situation, a full string theory description on the gravity side is not readily available due to intricacies of the $AdS_5 \times S^5$ sigma model, although checks of the correspondence (namely the spectra) have been performed and the presence of integrable structures have been proposed on both sides \cite{Frolov:2005ty,Frolov:2005dj,Frolov:2005iq}. In the latter, the string theory side is potentially easier to grasp, being related to the D1-D5 system, but the resulting deformation on the gauge theory side is somewhat peculiar to handle due to the warping of the near-horizon $AdS_3$ factor.

This paper aimed at addressing an intermediate situation in which a simple worldsheet description is available on the gravity side and tracking the deformation of the corresponding gauge theory is fairly straightforward.  The deformation of the compact part of the background we considered --- obtained as an exactly marginal \emph{asymmetric} deformation of the $SU(2) \times U(1)^4$ WZW worldsheet theory --- can be seen as an $AdS_3$/CFT${}_2$ counterpart of the Lunin-Maldacena deformation of $AdS_5 \times S^5$.\footnote{Though, strictly speaking, the equivalent of the Lunin-Maldacena deformation would have been a \emph{symmetric} deformation, preserving a $U(1) \times U(1)$ isometry rather than $SU(2) \times U(1)$.}  Similar to its higher-dimensional counterpart, the compact background resulting from deforming the $SU(2)\times U(1)$ WZW model can also be obtained as a TsT transformation that acts on the undeformed $S^3 \times T^4$ background (details on the relationship between TsT transformations and current-current deformations will be presented elsewhere).  Here, however, the origin of supersymmetry breaking in the spacetime dual gauge theory was seen directly from the spectrum of worldsheet operators in the deformed theory, and we identified a marginal operator in the dual theory that yields the same qualitative effect of supersymmetry breaking/enhancement as the marginal deformation of the worldsheet theory creates.

The same enhancement of supersymmetry was observed in warped $AdS_3$ spaces, and an additional peculiar enhancement of spacetime Virasoro algebras (from one to two) occurred at rational values of the deformation parameter $\frac{2\textsc{h}R}{\sqrt{k}}$ \cite{Detournay:2010rh}.  In recent work \cite{Hofman:2011zj}, it was shown that in a two-dimensional field theory with left global scaling symmetry there always exists a left conformal symmetry, along with either a right conformal symmetry or a left affine current algebra --- it would be interesting to understand whether there's any relation between these two observations.  In particular, understanding the latter case is of direct interest to the conjectured duality between warped $AdS_3$ spaces and a two-dimensional CFT \cite{Anninos:2008fx}, and also to the Kerr/CFT correspondence \cite{Guica:2008mu}.  A possible approach to clarify the situation would be to add probe branes to the deformed background since, generally, probe computations provide relevant information on the deformed dual gauge theory.\footnote{We thank E. Imeroni for pointing this out to us.}  In particular, in other contexts it appears to be a generic feature that additional D-brane probe configurations appear at specific rational values of the TsT deformation parameter, corresponding to new branches of vacua on the gauge theory side \cite{Imeroni:2008cr}.  It would be interesting to investigate the counterpart of this phenomenon in the D1-D5 or F1-NS5 system.

A more general and ambitious objective would be to relate worldsheet and spacetime CFTs in more general contexts, in particular when the worldsheet theory is not described by a WZW model.  For instance, the pure spinor formalism has made it possible to construct a quantizable sigma model for superstrings in an $AdS_5 \times S^5$ \cite{Berkovits:2000fe, Berkovits:2000yr,Vallilo:2002mh,Berkovits:2004xu,Berkovits:2007zk}, so string theory in the Lunin-Maldacena background might be viewable as an exactly marginal deformation of the $AdS_5 \times S^5$ worldsheet theory \cite{Frolov:2005ty}.  The usual approaches, however, deal with semi-classical strings and don't identify the marginal worldsheet operators responsible for the deformation. 
A potential intermediate step might be to consider strings on $AdS_3 \times S^3$ with R-R fluxes, building on \cite{Ashok:2009jw}, by studying the possible marginal deformations of the worldsheet CFT that could deform the background to $\mathit{WAdS}_3 \times S^3$ or $AdS_3 \times \mathit{WS}^3$, still supported by R-R fluxes rather than NS-NS fluxes.

\section*{Acknowledgments}

We would like to thank
S. Avery,
G. Comp\`ere,
M. Henneaux,
E. Imeroni,
D. Isra\"{e}l,
M. Petropoulos,
A. Strominger,
J. Troost,
and A. Virmani,
for various useful discussions and correspondences, and we'd especially like to thank D. Isra\"{e}l for helpful feedback throughout this work and on drafts of this paper.  We would also like to thank l'Institut d'Astrophysique de Paris, l'\'{E}cole Polytechnique, l'Universit\'{e} Libre de Bruxelles, and McGill University, for their hospitalities while this work was being completed. We also thank Opmoc for inspirational exchanges while this work was being finalized. The work of S.D. is funded in part by the European Commission through the grant PIOF-GA-2008-219950 within the FP7 Marie Curie Program.  The work of J.L. was funded in part by the Kavli Institute for Theoretical Physics, by NSF Grant PHY05-51164, by NSF Grant PHY07-57035, and by the National Science and Engineering Research Council of Canada.  The work of M.R. was funded in part by DOE Grant DE-FG02-91ER40618.

\appendix

\section{Representations of $SO(2,2)$}\label{app:reps}

In the $SL(2;\mathbb{R})$ WZW model approach, representations of $\mathfrak{so}(2,2) \cong \mathfrak{sl}(2;\mathbb{R}) \oplus \mathfrak{sl}(2;\mathbb{R})$ are written in terms of $SL(2;\mathbb{R})^2$ quantum numbers.  This is slightly inconvenient for understanding the relation with spacetime fields of definite spin since the transformation properties of such a field are defined in terms of representations of the Lorentz subgroup $SO(1,2) \subset SO(2,2)$.  It is therefore important to understand how to relate $SO(1,2)$ quantum numbers to those of $SL(2;\mathbb{R})^2$.

We can recast the $SO(2,2)$ algebra by realizing $AdS_3$ as a hyperboloid in $\mathbb{R}^{2,2}$.  The generators are then given in terms of the usual $(2+2)$-dimensional Lorentz generators $M_{ab}$ and $M_{a4}$, where  $a,b,\ldots = 1,2,3$. The metric in $\mathbb{R}^{2,2}$ is taken to be $\eta_{mn} = \mathrm{diag}(1,1,-1,-1)$ where $m,n,\ldots = 1,2,3,4$.  Defining
\begin{eqnarray}
L_1 \equiv M_{31} \, , &\!\!\!\!\!\!& \qquad L_2 \equiv M_{32} \, ,  \qquad  L_3 \equiv M_{12} \, ,  \\
R_1 \equiv -M_{24} \, ,  &\!\!\!\!\!\!& \qquad R_2 \equiv M_{14} \, ,  \qquad  R_3 \equiv M_{34} \, ,
\end{eqnarray}
we have the commutation relations
\begin{eqnarray}
[L_a,L_b] = i\epsilon_{ab}^{\phantom{ab}c} L_c \, , \qquad [L_a, R_b] = i\epsilon_{ab}^{\phantom{ab}c} R_c \, ,  \qquad  [R_a, R_b] = i\epsilon_{ab}^{\phantom{ab}c}L_c \, ,
\end{eqnarray}
where $\epsilon_{ab}^{\ph{ab}c}$ are the $\mathfrak{sl}(2;\mathbb{R})$ structure coefficients.  Note that the $AdS$ scale is hidden in $\mathbf{R}$ via $\mathbf{R}^2 = \frac{1}{\Lambda}\nabla^2$.
Then the generators of $\mathfrak{sl}(2;\mathbb{R})^{\oplus 2}$ are simply
\begin{equation}
\mathbf{J} = \tfrac{1}{2}\big( \mathbf{L} + \mathbf{R} \big) \, ,  \qquad  \bar{\mathbf{J}} = \tfrac{1}{2}\big( \mathbf{L} - \mathbf{R} \big) \, .
\end{equation}

The canonical quantum numbers to choose for an $SL(2;\mathbb{R})$ WZW model are $\mathbf{J}^2,\, J_3,\, \mathbf{\bar{J}}^2,$ and $\bar{J}_3$.  However, since we are interested in writing a vertex operator corresponding to a vector field in spacetime, we want to use a set of quantum numbers containing $\mathbf{L}^2$ and $L_3$.  It turns out that a maximal set is then $\mathbf{L}^2,\, L_3,\, \mathbf{R}^2,$ and $\mathbf{L}\cdot \mathbf{R}$.  Since we are interested in a vector representation of the Lorentz subgroup $SO(1,2)$, we want a three-dimensional representation.  Choosing the $L_a$ to be Hermitian, there exists a spin 1 representation, $l=1$ ($\mathbf{L}^2=-2l(2l-1)=-2$), though it is not always mentioned in constructions of $SL(2;\mathbb{R})$ representations because it contains a negative norm state (as it should, since $\eta^{33}A_3 A_3 \leq 0$).

Setting $l=1$ and $\hat{\mu}=-1,0,1$, therefore corresponds to a field that transforms as a spacetime vector.  We also have to specify $\mathbf{R}^2$ and $\mathbf{L}\cdot\mathbf{R}$ quantum numbers, which we will call $\mathfrak{r}$ and $\mathfrak{s}$, respectively, so that we can decompose
\begin{equation}
\label{eqn:so12tosl2}
\mathcal{U}_{l=1,\hat{\mu},\mathfrak{r},\mathfrak{b}} = \sum_{j,\bar{j},m,\bar{m}} c^{l=1,\hat{\mu},\mathfrak{r},\mathfrak{s}}_{j\bar{j}m\bar{m}} \,\mathcal{V}_{j\bar{j}m\bar{m}} \, .
\end{equation}
The relations among the generators yield
\begin{itemize}
\item $~~m+\bar{m}=\hat{\mu}$ ,
\item $~~-2l(2l-1)+\mathfrak{r} = \mathfrak{r}-2 = -2j(j-1)-2\bar{j}(\bar{j}-1)$ ,
\item $~~\mathfrak{s} = -j(j-1)+\bar{j}(\bar{j}-1)$ ,
\item $~~\mathbf{L}^2 - \mathbf{R}^2 = 4 \mathbf{J}\cdot\mathbf{\tilde{J}}$ ,
\end{itemize}
The final condition is equivalent to a constraint on the Clebsch-Gordan coefficients
\begin{equation}
\Big(-l(l-1)-\mathfrak{r}+8m\bar{m}\Big) c^{1\hat{\mu}\mathfrak{r}\mathfrak{s}}_{j\bar{j}m\bar{m}} = 2\Big(c^{1\hat{\mu}\mathfrak{r}\mathfrak{s}}_{j\bar{j},m-1,\bar{m}+1}(j-m+1)(\bar{j}+\bar{m}+1)+ c^{1\hat{\mu}\mathfrak{r}\mathfrak{s}}_{j\bar{j},m+1,\bar{m}-1}(j-m-1)(\bar{j}+\bar{m}-1)\Big) \, .\\
\end{equation}
In $SL(2;\mathbb{R})$ WZW models, we set $j=\bar{j}$ (this is important for level-matching), so let us apply this and set $\mathfrak{s}=0$.  Then we have $m+\bar{m}=\hat{\mu}$ and
\begin{equation}
j(j-1) = \frac{2-\mathfrak{r}}{4} \qquad \Longrightarrow \qquad  j = \frac{1}{2} \pm \frac{1}{2}\sqrt{3-\mathfrak{r}}  \ .
\end{equation}

\section{Affine $SU(2)$ Level from Asymptotic Symmetries}

The asymptotic symmetry algebra of a background is generated from `large gauge transformations'.  Often they contain exact symmetries of the background, such as Killing vectors of the theory.  To be a bit more concrete, the gauge symmetry (reducibility) parameters of the NS-NS fields of string theory are given by vectors $\xi$ and one-forms $\lambda$ satisfying
\be  \nabla_\xi \Phi = 0 \, ,  \qquad  \cL_\xi g = 0\, , \qquad     \cL_\xi B + d\lambda = 0 \, .
\ee
Asymptotic symmetry parameters solve these equations for large $r$ only and are associated with finite, integrable and asymptotically conserved charges (see appendix \ref{app:ASG} for more details).  They form a Lie algebra under the bracket\footnote{Note that  the one-form gauge parameters $\lambda$ are only defined up to a gauge transformation $\lambda \ra \lambda + d \phi$ for some function $\phi$. Under such a gauge transformation, $[\lambda, \lambda'] \rightarrow[\lambda + d \phi, \lambda' + d\phi' ] = [\lambda, \lambda'] + d \left(\mathcal{L}_\xi \phi' - \mathcal{L}_{\xi'} \phi\right)$. It is therefore enough that the commutation relations be satisfied up to an exact one-form.}
\be
\Big[ \big(\xi',\lambda'\big) , \big(\xi,\lambda\big) \Big] = \Big( [\xi',\xi] \, ,  \, 
[\lambda',\lambda] \Big) \, ,
\ee
where $[\lambda',\lambda]\equiv\mathcal{L}_{\xi'} \lambda - \mathcal{L}_\xi \lambda'$.
The asymptotically conserved charges will satisfy a Lie algebra similar to that of the asymptotic symmetry parameters, but it can also contain central terms such as the Brown-Henneaux central charge of $AdS_3$ \cite{Brown:1986nw}.

\subsection{$S^3$ Affine Level}
\label{app:affineSU2}

The $S^3$ with metric $ds^2(S^3) \propto \left(d\alpha^2+d\beta^2+d\gamma^2+2\cos\beta \, d\alpha d\gamma\right)$ has Killing vectors given by
\bea
 k^1 &=& \csc \beta \sin \gamma \; \p_\alpha    + \cos \gamma  \; \p_\beta     - \cot \beta \sin \gamma \; \p_\gamma  \, , \nonumber  \\
 k^2 &=&  \csc \beta \cos \gamma \; \p_\alpha    - \sin \gamma  \;   \p_\beta     -  \cot \beta \cos \gamma \; \p_\gamma  \, ,    \\
 k^3 &=& \p_\gamma  \, , \nonumber
 \eea
with similar expressions for the other $SU(2)$ factor. The gauge parameters are given by $\kappa^a = -\underline{k}^a - i_{k^a} B$  ($ \underline{k}^a$ are the 1-forms dual to the Killing vector fields, obtained by lowering an index with the string-frame metric):
\bea
 \kappa^1 &=&  \frac{k}{4}\Big(\csc \beta  \sin \gamma d\alpha  + \cos \gamma d\beta  + \cot \beta  \sin \gamma d\gamma \Big) \, , \nonumber \\
 \kappa^2 &=&  \frac{k}{4}\Big( \csc \beta  \cos \gamma d\alpha  - \sin \gamma d\beta  + \cot \beta  \cos \gamma d\gamma \Big)  \, ,   \\
 \kappa^3 &=&  \frac{k}{4}\, d\gamma \, .  \nonumber
\eea

While the global symmetry algebra is $SU(2)\times SU(2)$, we know that the algebra is enhanced to an affine $\widehat{SU(2)}\times\widehat{SU(2)}$.  Using the results of \cite{Giveon:1998ns,Kutasov:1999xu}, one can 
% One expects the exact symmetries to be promoted to an infinite-dimensional asymptotic symmetry algebra of the form
% \bea \label{affine}
%   [K^a_n, K^b_m] = \varepsilon^{ab}_{\;\;\;c}  K^c_{n+m}, \quad [L_n, K^a_m] = i m K^a_{n+m},
%  \eea
% for both sets of generators, with holomorphic generators commuting with the antiholomorphic ones. These commutation relations actually fix the form of the generators $K^a_n = (k^a_n, \kappa^a_n)$ of the affine extension almost uniquely. One gets 
% \bea
%  k^a_n &=& e^{i n (t+\theta)} k^a_0, \quad  \bar{k}^a_n = e^{i n (t-\theta)} k^a_0, \nonumber\\
%   \kappa^a_n &=& e^{i n (t+\theta)} \kappa^a_0, \quad  \bar{\kappa}^a_n = e^{i n (t-\theta)} \kappa^a_0, \quad a=1,2,3.
% \eea
explicitly determine the asymptotic symmetry parameters and compute the central extensions. Working in global coordinates on $AdS_3$, we can write the metric as
\be
ds^2(AdS_3) \propto -\frac{dt^2}{1-\rho^2} + \frac{d\rho^2}{(1-\rho^2)^2} + \frac{\rho^2}{1-\rho^2}d\theta^2 \, ,
\ee
where $0\leq\rho<1$ and $\rho\rightarrow 1$ is the boundary ($\rho$ is related to a more common radial coordinate $r$ through $r\equiv\frac{\rho}{\sqrt{1-\rho^2}}$), we have the $SL(2;\mathbb{R})\times SL(2;\mathbb{R})$ Killing vectors
\bea
\label{eqn:killing-global1}
J^- = \tfrac{1}{2}e^{-i(t+\theta)} \Big(-\rho\p_t -i (1-\rho^2)\p_\rho - \tfrac{1}{\rho} \p_\theta \Big)\, ,     &  \quad  &        \bar{J}^- =  \tfrac{1}{2}e^{-i(t-\theta)} \Big(-\rho\p_t -i (1-\rho^2)\p_\rho + \tfrac{1}{\rho} \p_\theta \Big)\, ,   \nonumber \\
J^3 = -\tfrac{i}{2} \big( \p_t + \p_\theta \big) \, ,   & \quad & \bar{J}^3 = -\tfrac{i}{2}\big( \p_t - \p_\theta\big)\, ,  \\
J^+ = \tfrac{1}{2}e^{-i(t+\theta)} \Big(\rho\p_t -i (1-\rho^2)\p_\rho + \tfrac{1}{\rho} \p_\theta \Big)\, ,    &  \quad  &   \bar{J}^+ = \tfrac{1}{2}e^{-i(t+\theta)} \Big(\rho\p_t -i (1-\rho^2)\p_\rho - \tfrac{1}{\rho} \p_\theta \Big)\, . \nonumber
\eea

In \cite{Kutasov:1999xu}, Kutasov and Seiberg wrote the spacetime affine $\widehat{SU(2)}$ generators as
\be
K^a(x,\bar{x}) = -\frac{1}{\pi} \int d^2z\, k^a(z,\bar{z}) \p_{\bar{z}}\Lambda(x,\bar{x};z,\bar{z}) \, ,
\ee
where $\Lambda(y,\bar{y};z,\bar{z})$ is defined through its commutation relations with the spacetime $SL(2;\bb{R})^2$ Killing vectors
\begin{align}
& [ J^-, \Lambda] = -\p_y\Lambda \, ,   & [\bar{J}^-, \Lambda ] &= -\p_{\bar{y}}\Lambda  \, , \nonumber \\
&\lbrack J^3 , \Lambda ] = -(y\p_y + 1)\Lambda  \, ,  & [\bar{J}^3 , \Lambda ] &= -\bar{y}\p_{\bar{y}} \Lambda \, , \\
&\lbrack J^+, \Lambda] = 1- (y^2\p_y + 2y) \Lambda \, ,   & [\bar{J}^+, \Lambda] &= -\bar{y}^2\p_{\bar{y}}\Lambda \, , \nonumber
\end{align}
and $\Lambda$ depends on $(z,\bar{z})$ through the spacetime coordinates $(t,\rho,\theta)$, which one should now think of as worldsheet scalar fields.  Using the form of the Killing vectors in global coordinates (\ref{eqn:killing-global1}), this determines
\be
\label{eqn:lambda}
\Lambda(x,\bar{x};z,\bar{z}) = \frac{(\bar{x}-i \rho e^{i(t-\theta)})}{(x-i \rho e^{i(t+\theta)})(\bar{x}-i\rho e^{i(t-\theta)})-(1-\rho^2) e^{2it}} \ .
\ee
$\Lambda$ does not transform as a tensor in the dual theory, but $\p_{\bar{z}}\Lambda$ does, so we can perform a decomposition into modes of $K^a(x,\bar{x})$.  As explained in \cite{Kutasov:1999xu}, this quantity is holomorphic and has dual conformal dimension $(1,0)$, so we have the modes 
\bea
K^a_{m} &=& -\frac{1}{2\pi^2 i} \oint dx\, x^{m} \int d^2z \, k^a(z,\bar{z}) \frac{-e^{i(t+\theta)}}{\rho\left(x-\frac{i}{\rho} e^{i(t+\theta)}\right)^2}\Big( i\p_{\bar{z}}\ln(\rho) + \p_{\bar{z}}(t+\theta) \Big)  \nonumber \\
&=& \frac{m}{\pi} \int d^2z \, k^a(z,\bar{z}) \Big( \frac{i}{\rho} e^{i(t+\theta)} \Big)^{m-1} \frac{e^{i(t+\theta)}}{\rho}\Big( i\p_{\bar{z}}\ln(\rho) + \p_{\bar{z}}(t+\theta) \Big)  \nonumber \\
&=&  -\frac{1}{\pi} \int d^2z \, k^a(z,\bar{z})\p_{\bar{z}} \Big(\frac{i}{\rho}e^{i(t+\theta)} \Big)^{m}  \, .
\eea
Defining $\Gamma \equiv \frac{i}{\rho} e^{i(t+\theta)}$, we can write this as
\be
K^a_m = \oint \frac{dz}{2\pi i} k^a(z,\bar{z}) \Gamma^m \, ,
\ee
so these are normalized correctly to form the spacetime affine $\widehat{SU(2)}$.

When we deform the sigma model action by adding one of these operators,
\be
S' = \frac{1}{4\pi} \int d^2 z \bigg\{ \big( G_{MN} + B_{MN} \big) \p X^M \bp X^N \bigg\} + K^a_m \, ,
\ee
it induces a combined transformation
\be
(G,B) \rightarrow (G,B) - 2\big( \underline{k}^a \otimes_s d\Gamma^m ,\, \underline{k}^a \wedge d\Gamma^m \big) \, ,
\ee
where $\underline{k}^a$ is the one-form obtained from the Killing vector $k^a$ by lowering with the string-frame metric. We define $k^{a}_{m}$ as the vector field that induces this transformation on $G$, and $\kappa^{a}_{m}$ as the one-form gauge transformation for $B$. They are defined through the equations
\be\label{transfGB}
\delta G=-2\mathcal{L}_{k^{a}_{m}}G \, , \qquad \delta B=-2(\mathcal{L}_{k^{a}_{m}}B+d\kappa^{a}_{m}) \, .
\ee
The metric deformation comes from a large diffeomorphism, and the vector field $k^{a}_{m}$ can be computed, being given by
\be
-2k^a_m = -2\Gamma^m k^a \, .
\ee
The transformation of the $B$-field is a combination of this diffeomorphism and a gauge transformation. Equation (\ref{transfGB}) defines $\kappa^a_m$ up to addition of a closed one-form.  Guessing $\kappa^a_m = \Gamma^m \kappa^a$, we can write this as
\bea
\underline{k}^a \wedge d\Gamma^m &=& d\iota_{k^a_m} B + \iota_{k^a_m} H + \Gamma^m d\kappa^a - \kappa^a \wedge d\Gamma^m   \nonumber \\
&=& \Gamma^m \big( \mathcal{L}_{K^a} B + d\lambda^a \big) - \big( \iota_{K^a} B + \lambda^a \big) \wedge d\Gamma^m \, ,
\eea
suggesting
\be
\kappa^a = -\underline{k}^a -  \iota_{k^a} B   \qquad\textrm{and}\qquad d\kappa^a = -\mathcal{L}_{k^a} B \, ,
\ee
which is consistent since $d\iota_{k^a} B = \mathcal{L}_{k^a} B - \iota_{k^a} H_3$ and $d\underline{k}^a = \iota_{k^a} H_3$ (noting that $H_3\propto f_{abc} \underline{k}^a \wedge \underline{k}^b \wedge \underline{k}^c$ and $d\underline{k}^a \propto f^a_{\ph{a}bc} \underline{k}^b \wedge \underline{k}^c$).  So we have the asymptotic reducibility parameters
\be
K^a_m~ \longleftrightarrow~ \big( k^a_m , \kappa^a_m \big) = \Gamma^m \big( k^a , \, -\underline{k}^a - \iota_{k^a} B \big) \, .
\ee
Following appendix \ref{app:ASG}, we expect a central term to appear in the algebra, and its value $k_{\mathrm{st}}$ will arise from computing
\be
\int \mathbf{k}_{K^a_m}[\delta_{K^b_m} (g,B), (g,B)] \, ,
\ee
with the charge one-form given by (\ref{eqn:charge-one-form}). The gravitational contribution $\mathbf{k}^g_{K^a_m}$ can be shown to vanish. The $B$-field contribution is obtained from \re{Bcharge} as
\begin{align}
&\frac{1}{16\pi G_6}\int_{S_{AdS_3}^1 \times S^3} \bigg\{ \iota_{
k^b_n} \Big( e^{-\Phi} \delta_{K^a_m} B \wedge \star H_3 \Big) - e^{-
\Phi} \Big( \iota_{k^b_n} \delta_{K^a_m} B \wedge \star H_3 + \big( 
\iota_{k^b_n} B + \kappa^b_n \big) \wedge \mathcal{L}_{k^a_m} 
\big(\!\star\!H_3) \Big)\bigg\} \nonumber \\
& = \frac{1}{16\pi G_6}\int_{S_{AdS_3}^1 \times S^3} \bigg\{ e^{-\Phi} \delta_{K^a_m} B \wedge \iota_{k^b_n} \big( \!\star \!H_3\big) - e^{-\Phi} \big(\iota_{k^b_n} B + \kappa^b_n \big) \wedge d \iota_{k^a_m}\big(\!\star\! H_3\big) \bigg\} \, ,
\end{align}
where $S^1_{AdS_3}$ refers to the AdS$_3$ boundary (parameterized by $\theta$) and $G_6^{-1} = g_s^{-2} \textrm{Vol}(T^4) G_{10}^{-1}$.  Noting that $\big(\iota_{k^b_n} B + \kappa^b_n \big) = \Gamma^n \big( \iota_{k^b} B + \lambda^b\big) = -\Gamma^n \underline{k}^b$  ($\Gamma^n \sim e^{i n (t+\theta)}$), this becomes
\begin{align}
\frac{1}{16\pi G_6}\int_{S_{AdS_3}^1 \times S^3} e^{-\Phi}\bigg\{  \underline{k}^a \wedge \Gamma^n d\Gamma^m \wedge \iota_{k^b} \big( \!\star \!H_3\big) -\underline{k}^b \wedge \Gamma^m  d \Gamma^n \wedge \iota_{k^a}\big(\!\star\! H_3\big) \bigg\} \, ,
\end{align}
where we also used the fact that $d\underline{k}^b \wedge \iota_{k^a}\big(\!\star\! H_3\big)$ is a 4-form on $S^3$ and therefore vanishes.  Similarly, we can conclude that $\int_{S^3} \iota_{k^a} \big( \underline{k}^b \wedge \star H_3\big) =0$, so we have
\begin{align}
\frac{1}{16\pi G_6}\int_{S_{AdS_3}^1 \times S^3}   (n-m) \iota_{k^b} \underline{k}^a \Big\{ \Gamma^{m+n-1} d\Gamma \wedge e^{-\Phi}\!\star \!H_3 \Big\} \, ,
\end{align}
where we used the fact that $\iota_{k^a}\underline{k}^b = \frac{k}{4}\delta^{ab}$ ($\underline{k}^a$ is defined from $k^a$ using the string-frame metric, whereas the Einstein frame metric is the one appearing elsewhere).  When $m+n\neq 0$, the integrand is periodic in $\theta$, so the integral vanishes.  When $m+n=0$, the integrands contains a $\theta$-independent term linear in $d\theta$, which makes a nonvanishing contribution:
\be
-\frac{k}{16G_6} m\delta^{ab}\delta_{m+n,0} \int_{S^3}  e^{-\Phi} \star \!H_3 = \frac{k}{16 G_6} m\delta^{ab}\delta_{m+n,0} (2k) (2\pi^2) e^{-\Phi} = \frac{\left(2\pi^2 k^{\frac{3}{2}} e^{-\frac{3}{4}\Phi}\right)}{4G_6} \frac{\sqrt{k}\,e^{-\frac{1}{4}\Phi}}{2} m \delta^{ab}\delta_{m+n,0} \, .
\ee

Since the Einstein-frame volume of $S^3$ is $2\pi^2 k^{\frac{3}{2}}e^{-\frac{3}{4}\Phi}$, the $AdS_3$ radius is $\ell^2=k e^{-\frac{1}{2}\Phi}$, and the central term should be $\frac{k_{\textrm{st}}}{2}m\delta^{ab}\delta_{m+n,0}$, the level is seen to be
\be
k_{\textrm{st}} = \frac{\sqrt{k}\, e^{-\frac{1}{4}\Phi}}{4 G_3} = \frac{\ell}{4 G_3}
\ee
since $G_3^{-1} = G_6^{-1} \textrm{Vol}(S^3)$.  Thus, the level of the spacetime affine $\widehat{SU(2)}$ is consistent with the two-dimensional $\mathcal{N}=(4,4)$ superconformal algebra which demands the level be $\frac{c_{\textrm{st}}}{6}$. It is interesting to observe that while the Brown-Henneaux central charge came exclusively from the gravitational sector of the theory, the affine level originates purely in the $B$-field gauge transformations contributions.

\subsection{Warped $S^3$ Affine Level}

From the coordinate transformation (\ref{eqn:coordxform}), we know that we will always locally have $SU(2)\times SU(2)$ Killing vectors, but that these will only respect the global identifications when $\frac{2\textsc{h}R}{\sqrt{k}}\in\mathbb{Z}$.  We can then only write down the spacetime affine $\widehat{SU(2)}\times \widehat{SU(2)}$ when this condition holds, and in those cases we will obtain the same result as above for the level of the spacetime affine $\widehat{SU(2)}\times \widehat{SU(2)}$ since it is precisely those cases in which the global structure of the spacetime is $S^3 \times S^1$, the case analyzed above.

%%%%%%%%%%%%%%%%%%%%%%%%  Charge computation %%%%%%%%%%%%%%%%%%

\def\oneone{\rlap 1\mkern4mu{\rm l}}
\def\beq{\begin{eqnarray}}
\def\eeq{\end{eqnarray}}

\section{Computation of Surface Charges}
\label{app:ASG}

This appendix reviews the formalism of \cite{Barnich:2001jy,Barnich:2003xg,Barnich:2007bf} which we use to compute asymptotically conserved charges for our ten-dimensional theory (see also \cite{Compere:2007az} and appendix A of \cite{Compere:2009qm}).
Our $D$-dimensional theory takes the generic form
\begin{equation}
I = \frac{1}{16 \pi G} \int \, \left( R \, \star {\oneone}
- \frac{1}{2}  \star d \chi \wedge d \chi
-  \frac{1}{2}  e^{\alpha .
\chi } \star \mathbf H \wedge \mathbf  H \right)
,\label{gaction}
\end{equation}
where $\chi$ is a scalar field and $\mathbf H$ is a three-form field strength. We will denote the set of fields by $\phi = (g, \mathbf B, \chi)$, where $\mathbf B$ is a two-form potential for $ \mathbf  H$.  Associated to every asymptotic Killing vector $\xi$,\footnote{Asymptotic Killing vectors are defined as diffeomorphisms that satisfy the Killing equations in an asymptotic region and are associated with finite, conserved, and integrable charges.} there is a space-time $D-2$ form
\begin{eqnarray}
\mathbf k_{\xi} [\delta \phi ; \phi ] \label{oneform}
\end{eqnarray}
that is linear in $\delta\phi$ and its derivatives --- it is a one-form in `field space'.  $\mathbf k_{\xi} [\delta \phi ; \phi ]$, which can be constructed by a well-defined algorithm that depends only on the equations of motion, is the basic ingredient in the definition of asymptotically conserved charges \cite{Barnich:2001jy,Barnich:2003xg,Barnich:2007bf} (a similar expression exists for any gauge symmetry parameter of the theory).
It enjoys the following properties:
\begin{itemize}
\item  Given a solution to the equations of motion, $\tilde{\phi}$, and a variation $\delta\phi$ that satisfies the linearized equations of motion around $\phi=\tilde{\phi}$, then for every exact Killing vector $\xi$ of the background $\tilde{\phi}$, there exists a conserved quantity
\beq
\delta Q_{\xi} \equiv \oint_S \mathbf k_{ \xi} [\delta \phi ; \tilde\phi ]  \label{infcharge}
\eeq
that only depend on the homology class of the $(D-2)$-surface $S$.  $\delta Q_\xi$ defines the difference in charge between the backgrounds $\tilde{\phi}$ and $\tilde{\phi}+\delta\phi$ and is unique \cite{Barnich:1994db}.

\item When $\xi$ is an asymptotic Killing vector, the difference in charge between the solutions $\tilde{\phi}$ and $\tilde{\phi}+\delta\phi$ is given by
\beq
\delta Q_{\xi} \equiv \underset{r \rightarrow \infty}{\text{lim}} \oint_{S^r} \mathbf k_{ \xi} [\delta \phi ; \tilde\phi ] \, . \label{infchargeasympt}
\eeq

\item Since $\mathbf k_{ \xi} [\delta \phi ; \tilde\phi ]$ is constructed purely of the equations of motion and solutions $\tilde{\phi}$ and $\tilde{\phi}+\delta\phi$, it does not depend on boundary terms in the action.

\item Since $\mathbf k_{ \xi} [\delta \phi ; \phi ]$ is a linear functional of the equations of motion, it can be expressed as a sum of terms arising from each contribution to the Lagrangian.

\item Given two solutions, $\bar \phi$ and $\tilde\phi$, in the same phase space, for each asymptotic Killing vector $\xi$, the difference in charge between $\bar\phi$ and $\tilde\phi$ is given by
\begin{equation}\label{finitecharge}
Q_{\xi}[\tilde\phi,\bar \phi] \equiv \lim_{r\rightarrow\infty} \oint_{S^r} \int_
\gamma \mathbf{k}_{\xi}[\delta \phi',\phi'] + N_{\xi}[\bar \phi]\, ,
\end{equation}
where $\gamma$ is a path in field space connecting $\bar\phi$ with $\tilde\phi$, $\delta\phi$ and its derivatives are a basis for the line element along $\gamma$, and $N_ {\xi}[\bar \phi]$ is an arbitrary normalization
constant.  Demanding that the charge be independent of the path $\gamma$ implies an integrability condition that restricts the field space of $\phi$ as well as the space of asymptotic Killing vectors.

\end{itemize}
Additional properties of the charge form \eqref{oneform} are discussed in \cite{Barnich:2004uw,Barnich:2006av}.

For the Lagrangian \eqref{gaction}, the contributions to the $(D-2)$-form can be split into four pieces:
\beq
\label{eqn:charge-one-form}
\mathbf k_{ \xi} [\delta \phi ; \phi ] &=&
\mathbf k^{g}_{ \xi}[\delta g;g]  +  e^{\alpha \chi} k^{\mathbf B}_{ \xi}[\delta \phi ; \phi ] +  \mathbf k^{\chi}_{ \xi}
[\delta \phi;\phi]   + \, \mathbf k^{\mathbf B \, suppl}_
{\xi}[\delta \phi;\phi]   \, . \label{k_totg}
\eeq
The gravitational contribution to the charge form is given by \cite{Abbott:1981ff,Barnich:2001jy}
\begin{eqnarray}
\mathbf k^{g}_{ \xi}[\delta g;g] &=& -\delta \mathbf Q^g_{ \xi} -i_{\xi}\mathbf \Theta^g[\delta g] -\mathbf
E^g_\cL[\cL_\xi g, \delta g]\, ,  \label{grav_contrib}
\end{eqnarray}
where
\begin{subequations}
\begin{eqnarray}
\mathbf Q^g_{\xi} &=&  \star \Big(\frac{1}{2} (D_\mu \xi_\nu-D_\nu \xi_\mu)
dx^\mu \wedge dx^\nu \Big)\, ,\label{Komar_term} \\
\mathbf \Theta^g[\delta g]&=&\star \Big(  (D^\sigma \delta
g_{\mu\sigma}-
g^{\alpha\beta} D_\mu \delta g_{\alpha\beta})\,dx^\mu\Big)\, ,\\
\mathbf E^g_\cL[\delta_2 g, \delta_1 g] &=& \star \Big( \frac{1}{2} \delta_1
g_{\mu\alpha} g^{\alpha\beta }\delta_2 g_{\beta\nu} dx^\mu \wedge
dx^\nu \big)\, .
\end{eqnarray}
\end{subequations}
The term \re{Komar_term} is known as the Komar $(D-2)$-form while $E^g_\cL$, which does not appear in the Iyer-Wald
formalism~\cite{Iyer:1994ys}, vanishes for exact Killing vectors but may be relevant for asymptotic symmetries. In \eqref{grav_contrib} above and \eqref{Bcharge} below, $\delta$ is an operator that acts only on the fields $\phi$, not on the asymptotic Killing vectors $\xi$.
The $p$-form contribution to the charge form (here $p=2$) is given by \cite{Compere:2007vx}
\begin{equation}
\mathbf k^{\mathbf B}_{ \xi}[\delta \phi ; \phi ]=-
\delta \mathbf Q^{\mathbf B}_{\xi} + i_\xi \mathbf
\Theta_{\mathbf B}-\mathbf E^{\mathbf B}_\cL[\cL_\xi \mathbf B,\delta \mathbf B] \, , \label{Bcharge}
\end{equation}
where
\begin{eqnarray}
& &\mathbf{Q}^{\mathbf B}_{\xi}  =  i_\xi \mathbf B
\wedge \star \mathbf H \label{QA}\, ,\qquad\qquad\qquad \mathbf \Theta^{\mathbf B} = \delta \mathbf B
\wedge \star \mathbf H\,,\label{ThetaA}\\
& &\mathbf E^{\mathbf B}_\cL[\delta_2 \mathbf B,\delta_1 \mathbf B] =  \star \big(\frac{1}{2}
\frac{1}{(p-1)!}\delta_1
\mathbf B_{\mu\alpha_1\cdots \alpha_{p-1}} \delta_2 \mathbf B_{\nu}^{\;\,\,\alpha_1\cdots \alpha_{p-1}} dx^\mu\wedge
dx^\nu \big)\, .
\end{eqnarray}
Finally, the last two terms are given by
\beq
\mathbf k^{\chi}_{ \xi}[\delta \phi;\phi] &=& i_\xi \big(\star (d
\chi \delta \chi  ) \big) \, , \\ % \hspace{1cm}\ , \\
\mathbf k^{\mathbf B \, suppl}_{\xi}[\delta \phi; \phi] &= &  \alpha \, \delta \chi \; e^{-
\alpha . \chi} {\cal Q}^{\mathbf B}_{\xi} \, .
\eeq

The next step is the representation of the algebra of asymptotic Killing vectors by the asymptotically conserved charges \re{finitecharge}. For this, we need to define a set of fields  $(\phi, \delta \phi)$ (the \emph{phase space} of the theory) and gauge parameters $\xi$ (the \emph{asymptotic symmetries}) such that the charges $Q_\xi[\phi,\bar \phi]$ are all finite, asymptotically conserved, and integrable for all $\phi$ and $\bar \phi$ in the phase space.
One can then show (modulo a technical assumption) that for any solutions $\phi$ and $\bar\phi$ in the
phase space, and for any asymptotic symmetries $\xi, \xi^\prime, \lambda^
\prime$, the Dirac bracket defined by
\begin{equation}
\left\{ Q_{\xi}[\phi,\bar \phi],
Q_{\xi^\prime}[\phi,\bar \phi] \right\} \equiv
\oint_{S^\infty} \mathbf{k}_{\xi}[\cL_{\xi^\prime}
\phi,;\phi] \label{poissonbracket}
\end{equation}
can be written as
\begin{equation}
\left\{ Q_{\xi}[\phi,\bar \phi],
Q_{\xi^\prime}[\phi,\bar \phi] \right\} =
Q_{[\xi,\xi^\prime]}
[\phi,\bar \phi] -
N_{[\xi,\xi^\prime]}
[\bar \phi] + K_{\xi,\xi^\prime}[\bar \phi]\, ,\label{formula}
\end{equation}
where \begin{eqnarray}
K_{\xi,\xi^\prime}[\bar
\phi] = \oint_{S^\infty} \mathbf{k}_{\xi}[\cL_{\xi^\prime}
\bar\phi;\bar \phi] \label{eq:cc}
\end{eqnarray}
is a central extension that is nontrivial only if it
cannot be reabsorbed into the normalization $N_{[\xi, \xi^\prime]}[\bar \phi]$. An important observation is that the central term can be computed from the data of a background only, independent from the definition of a phase space (and when the phase space is known, the result is independent of the choice of a background in the phase space).

\bibliographystyle{utphys}
\bibliography{master4}

\end{document}